\documentclass[aps,prb,preprint,groupedaddress,showpacs,showkeys]{revtex4}

\usepackage{graphicx}
\begin{document}
\title{Molecular Correlation Functions for Uniaxial Ellipsoids in the
Isotropic State}

\author{Cristiano De Michele}
\email{cristiano.demichele@phys.uniroma1.it}
\homepage{http://www.crs-soft.it}
\affiliation{Dipartimento di Fisica and INFM-CRS Soft, Universit\'a di
Roma ``La
Sapienza'', P.le Aldo Moro 2, 00185 Roma, Italy }
 
\author{Antonio Scala}
\affiliation{Dipartimento di Fisica and INFM-CRS SMC, Universit\'a  di
Roma ``La Sapienza'', P.le Aldo Moro 2, 00185 Roma, Italy }
 
\author{Rolf Schilling}
\affiliation{Johannes-Gutenberg-Universitat Mainz, D-55099 Mainz,
Germany}
 
\author{Francesco Sciortino}
\affiliation{Dipartimento di~Fisica and INFM-CRS Soft, Universit\'a di
Roma ``La Sapienza'', P.le Aldo Moro 2, 00185 Roma, Italy }
 
 
\date{\today}
 
\begin{abstract}
We perform event-driven molecular dynamics simulations of a system
composed by uniaxial hard ellipsoids for different values of the
aspect-ratio and packing fraction . We compare the molecular
orientational-dependent structure factors  previously calculated
within the Percus-Yevick approximation with the numerical results.
The agreement between theoretical and numerical results is rather
satisfactory.  We  also show that, for specific orientational
quantities, the molecular structure factors are sensitive to the
particle shape and can be used to
 distinguish  prolate from oblate ellipsoids.  A first-order
theoretical expansion  around the
 spherical shape and a geometrical analysis of the configurations
confirms and explains such an observation.
\end{abstract}
 
\pacs{61.20.Ja, 61.25.Em, 61.30.Cz, 61.20.Gy}
 
\maketitle

\section{Introduction}

The structure of {\it simple} liquids can be characterized by the
density-density correlator $g(r)$ or by its Fourier transform, the static
structure factor $S(q)$. Strong efforts have been made to
derive analytical tools for the calculation of these quantities, since 
$g(r)$  or $S(q)$ allow one to calculate several thermodynamical quantities,
e.g.~the equation of state.  Among the  most prominent theoretical approaches 
are the Percus-Yevick (PY), the hypernetted chain (HNC) and other more elaborate integral equations 
\cite{Hansen86}.  For models of simple liquids --- when the number
density $\rho$ is not too large --- integral equations provide
predictions for $g(r)$ and $S(q)$ which well reproduce the "exact" results
evaluated from experiments or simulations.

For {\it molecular} liquids, structural information becomes
more diverse due to the presence of the orientational degrees of
freedom and of their interactions with the translational ones.
Expansion of the angular dependent microscopic density with
respect to spherical harmonics and Wigner functions for linear and
arbitrary molecules, respectively, leads to a generalization of
$S(q)$ to tensorial correlators $S_{\lambda \lambda'} ({\bf q})$
(see section II. B). Several analytical approaches have been proposed to
calculate these correlation functions. The simplest one performs
{\it ``Ans\"atze''} which relates $S_{\lambda \lambda'} ({\bf q})$,
or the corresponding direct correlation function $c_{\lambda
\lambda'} ({\bf q})$, to $S(q)$ or $c(q)$, $q=|{\bf q}|$ of an
appropriately chosen related {\it simple} liquid \cite{A1,A1bis,A2,A3,A4,A4bis,A5,A6}. 
Similar to this is the ``Ansatz'' for $c_{\lambda
\lambda'} ({\bf q})$ based on the geometry of two molecules
\cite{A7,A8,A9}. The thermodynamical perturbation theory is a
systematic approach which uses a liquid system  of {\it isotropic}
particles as a reference and considers the deviation of the particles from {\it
sphericity} as a perturbation. Different types of expansions
exist, depending on the quantity which is expanded. For more
details the reader may consult Ref. \cite{GrayGubbins}. Finally, the integral
equation theories have been extended to molecular liquids \cite{GrayGubbins,Hansen86}.

As compared to simple liquids, the quality of the integral equations for anisotropic particles
has been less intensively investigated. Furthermore, comparisons 
have been mainly restricted to hard
ellipsoids of revolution and to the value of the coefficients
$g_{ll'm} (r)$ and $c_{ll'm} (r)$ of the expansion into rotational
invariants of, respectively, the pair distribution and direct
correlation function in real space. The quality of PY and
HNC-theory for a liquid of hard ellipsoids \cite{A10} has been
tested against molecular dynamic simulation data fifteen years ago
\cite{A11}. Ref. \cite{A10} reports such a comparison together with
earlier MC-results \cite{A12}, however, restricted to the center
of mass correlator $g_{\rm 000} (r)$. Satisfactory agreement has
been found for both approximation schemes. PY-theory seems to be
inferior to HNC-theory, because (i) the PY-correlators for {\it
oblate} ellipsoids deviate stronger than the HNC-correlation from
the corresponding MD-results and (ii) PY-theory does not yield an
indication for an isotropic-nematic phase transition, in contrast
to HNC-theory. The good quality of PY-theory for {\it prolate}
ellipsoids has been confirmed \cite{A13,A14}. Particularly it has
been demonstrated that reasonably good PY-predictions require that
terms up to $l_{\rm max}=6$ are taken into account
\cite{A14}. In variance with the observation of Ref. \cite{A11} it
has been recently found that PY-theory also yields an
isotropic-nematic phase transition \cite{letzlatz}.

Finally we mention that the direct correlation functions $c_{ll'm}(r)$ obtained from a MC-simulation \cite{A15} have been
compared with the ``Ans\"atze'' described in Refs. \cite{A1,A2,A3,A4,A4bis,A5,A6}. 
Some of them exhibit a satisfactory agreement with the
MC-results, particularly for large $r$-values. But in our opinion
the agreement is less good than the one found from integral equation
theory \cite{A11, A13, A14}.

The above short review on previous investigations of hard ellipsoids makes it
obvious that the debate on the orientational structural properties is not quite settled.
Therefore, we have applied a newly developed event-driven
MD-algorithm \cite{Rapaport} in order to calculate the static molecular
correlation functions of hard ellipsoids in $q$ space. In contrast to previous work described
above, we have chosen the tensorial correlators $S_{\lambda \lambda'}({\bf q})$, 
$\lambda =(l,m)$ (see sect. II. B). 
They have the advantage that they can directly be deduced from
neutron and synchrotron radiation  scattering experiments, at
least for $l \leq 2$. One of our main goals is to compare accurate numerical results with 
 PY-predictions from Ref. [8], 
obtained for the gaussian overlap model \cite{A16} and with a
truncation at $l_{\rm max}=8$, a value larger than the one previously used in
Ref. \cite{A14}. Besides this, we will interpret the peak
structure of these correlators, particularly of the non-diagonal
one, $S_{\rm 2000} (\bf q)$. We will demonstrate that the
qualitative $q$-dependence of this nondiagonal correlator allows us to
easily distinguish between oblate and prolate ellipsoids. The
application of first order thermodynamic perturbation theory for $S_{\rm 2000} ({\bf q})$
will provide support to this finding . Finally,
we want to check how far the prediction of an isotropic-nematic
phase transition \cite{A15} found from the growth of $S_{\rm
2020} ({\bf q} =0)$ is reproduced by our MD-simulation.

The outline of the manuscript is as follows. In the next section we
will describe the model and the tensorial functions $S_{\lambda
\lambda'} ({\bf q})$. Section III contains results concerning:
(i) a comparison for $S_{\lambda \lambda'} ({\bf q})$ from
PY-theory and MD-simulation, (ii) a geometrical interpretation
of the $q$-dependence of $S_{\rm 2000} ({\bf q})$ and (iii) the
first order perturbation theory for $S_{\lambda \lambda '} ({\bf
q})$. The final section IV contains our conclusions.

\section{Methods}
 
\subsection{Model}
 
We study a system composed of  $N=256$ or $2048$  uniaxial hard
ellipsoids, i.e. ellipsoids with revolution axis of length $a$ and
two other axes of identical length $b$ in a volume $V$.  The aspect
ratio is defined as $X_{0} \equiv a/b$, with $0<X_0<\infty$. The
volume of each particle is $\pi X_{0}b^{3}/6$ and thus  the packing
fraction $\phi \equiv \pi X_{0}b^{3}\rho/6$  (where $\rho=N/V$ is the
number density). We perform event driven molecular dynamics
simulations, with periodic boundary conditions, at several values of
$0.4<\phi<0.51$  and several $0.4<X_0<2.8 $ values. Distances are
measured in units of the axes  geometric mean $l\equiv
\sqrt[3]{ab^{2}}$. Ellipsoids have mass $m=1$  and a spherically
symmetric momentum of inertia, i.e. $I_{x}=I_{y}=I_{z}=2 m r^{2}/5$
with $r=\min\{a,b\}/2$.
 
The  event-driven molecular dynamics simulation\cite{Rapaport} for a
one-component hard-ellipsoids system (HES) used in this work is
described in details in Ref.\cite{DeMicheleScala}.
The prediction of events (collisions among molecules) is different
from what has been proposed in the past. It relies on evaluations of
distance between objects based on gradient descent and Newton-Raphson
root finding algorithms. Such strategy in principle works for any
objects whose surface is differentiable and hence it is not limited to
ellipsoids. The efficiency of the algorithm  is comparable to the
algorithm recently proposed by  Donev-Torquato-Stillinger\cite{DonevTorqStill}.
 
\begin{figure}
\includegraphics[width=0.5\textwidth,keepaspectratio]{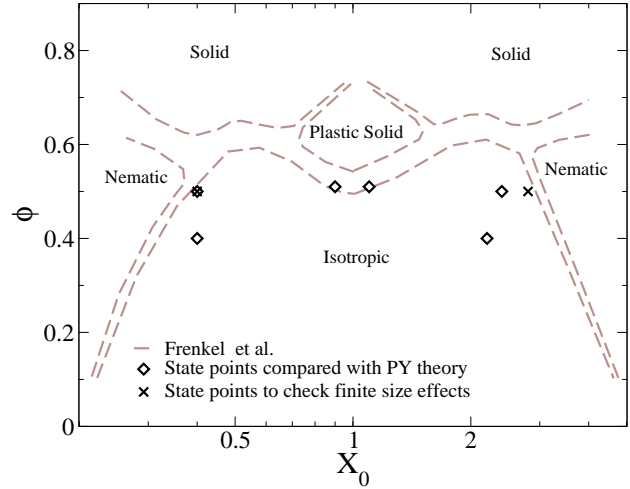}
\caption{Hard ellipsoids phase diagram. The dashed line are the phase
boundaries  calculated by \cite{FrenkelPhaseDiag}.
The open diamonds correspond to the points of the phase diagram which we compare with the PY results of\cite{letzlatz}. The {\cal X}'s correspond to the points of the phase diagram for which we have analized finite size effects}
\label{fig:Points_in_the_phase_diagram}
\end{figure}

To create the starting configuration,  we generate a random
configuration at very low $\phi$ and  grow independently the
particles up to the desired $\phi$ (i.e. we perform a quench in
$\phi$ at fixed $N$,$X_{0}$). To equilibrate the systems, we
simulate until the angular second Legendre polynomial correlation
function, for the angle associated to the axis of symmetry,  has
decayed to zero. In addition we check that particles have diffused
more than $\max\{a,b\}$. Simulations last approximatively from
$10^8$ to $10^9$ hard-ellipsoid collisions; between $1000$ and
$5000$ equally spaced configurations have been stored for the analysis. 
Fig.~\ref{fig:Points_in_the_phase_diagram}  shows the studied state points
together with the known thermodynamic lines. The majority of the
studied points lies close to the equilibrium transition line to
maximize the structural signatures in the static correlation
functions.

\subsection{Molecular structure factors}
 
A system of $N$ rigid molecules can be described by  the positions
of the centers  ${\bf r}_{j}$ and the orientation (Euler angles)
${\bf \Omega}_{j}$ of the $j$-th molecule. The microscopic density
$\rho\left({\bf r},{\bf \Omega}\right)=\sum_{j}\delta\left({\bf
r}-{\bf r}_{j}\right)\delta\left({\bf \Omega}-{\bf
\Omega}_{j}\right)$ can be expanded with respect to plane waves
$e^{i{\bf q}\cdot{\bf r}}$ and to Wigner matrices
$D_{mn}^{l*}\left({\bf \Omega}\right)$
\cite{GrayGubbins},\cite{water}. For molecules with a rotational
symmetry axis, ${\bf \Omega} \equiv (\theta,\phi)$, Wigner
matrices reduce to spherical harmonics and the microscopic density
can be expanded into tensorial modes
 
\[
\rho_{lm}\left({\bf q}\right)=\sqrt{4\pi}i^{l}{\displaystyle
\sum_{j=1}^{N}}e^{i{\bf q}\cdot{\bf r}_j}Y_{lm}\left({\bf
\Omega_j}\right)\]
 
\noindent where $l$ takes integer values $\geq 0$ and $m$ runs
between $-l$ and $l$. The factor in front of the sum is for
technical convenience. The molecular structure factors are defined
as
 
\[
S_{lml'm'}\left({\bf q}\right)=\frac{1}{N}\left\langle
\rho_{lm}^{*}\left({\bf q}\right)\rho_{l'm'}\left({\bf q}\right)\right
\rangle \]
 
\noindent that will in general depend both on the modulus and on
the orientation of ${\bf q}$. The symbol  $<....>$  indicates an
ensemble average. A convenient reference system is provided by the
$q$-frame, where the direction of the $z$-axis is parallel to
${\bf q}$ \cite{GrayGubbins}.  For molecules with a rotational
symmetry axis, molecular structure factors in the $q$-frame become
diagonal in $m$ \cite{SchillingScheidsteger} so that
$S_{lml'm'}=\delta_{mm'}S_{lml'm}$. For ellipsoids, the $S_{lml'm}
(q) $ have been calculated within the Percus-Yevick
approximation\cite{letzlatz} and used as input of mode-coupling
theory calculations to evaluate the glass transition lines
\cite{LetzSchilLatz}.  
%
\section{Results and discussion}
\subsection{Comparison between PY and simulation data for
$S_{lml'm'}$}
 
\begin{figure}
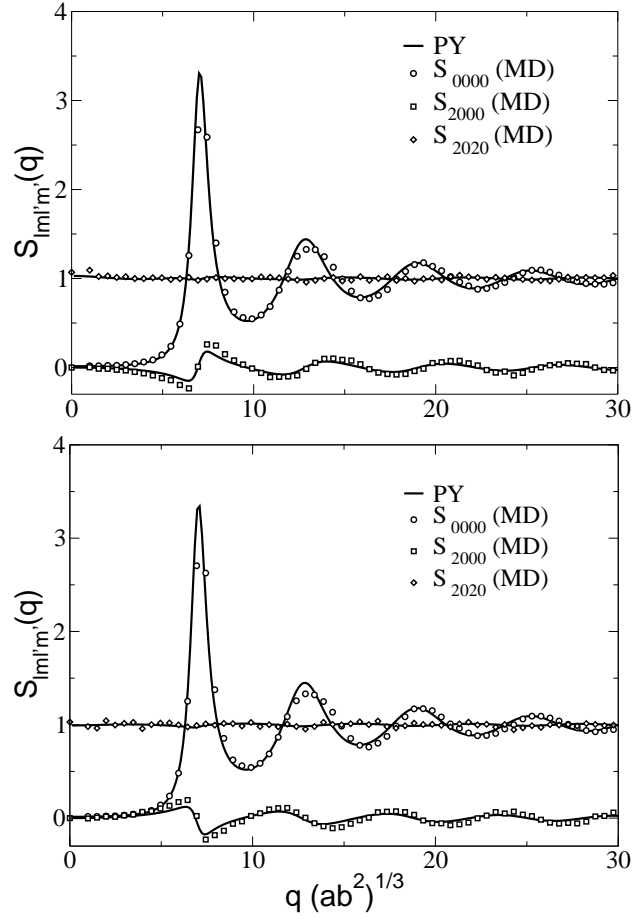

\includegraphics[width=0.5\textwidth]{PY-MD-Phi051EL09}
\vskip 0.2cm
\includegraphics[width=0.5\textwidth]{PY-MD-Phi051EL11}
\caption{$S_{lml'm'}\left({\bf q}\right)$ for  $X_{0}=0.9$ (top) and
$X_0=1.1$ (bottom) at $\phi=0.51$, i.e.  near the hard-sphere case
$X_{0}=1$. Symbols are
simulation results, lines are PY predictions from Ref.\protect
\cite{letzlatz}.
}
\label{fig:almost_hard_sphere}
\end{figure}

Fig.~\ref{fig:almost_hard_sphere}  compares the numerical results
and the PY predictions from Ref.\cite{letzlatz} for $S_{0000}$,
$S_{2000}$ and $S_{2020}$ when $X_0 \approx 1$ (an almost
hard-sphere case)  for both oblate and prolate ellipsoids.   In
all cases, the PY predictions satisfactorily describe the
numerical results. We note that $S_{0000}$ resembles the typical
shape of the HS fluid and is practically the same for oblate
($X_0=0.9)$ and prolate $X_0=1.1)$ ellipsoids whereas $S_{2020}$
is structureless for all $q$ values (as expected since the studied
points are far from the nematic phase). Nevertheless, information
on the angular structure is contained in $S_{2000}$. This function
shows an interesting $q$ behavior when comparing the prolate and
oblate case. Indeed, it appears that the prolate $S_{2000}(q)$ has
opposite sign as compared to the oblate one. We also notice that
the location of the first peak of $S_{0000}$ coincides with the
location of one extremum also in the $S_{2000}(q)$.
 
\begin{figure}
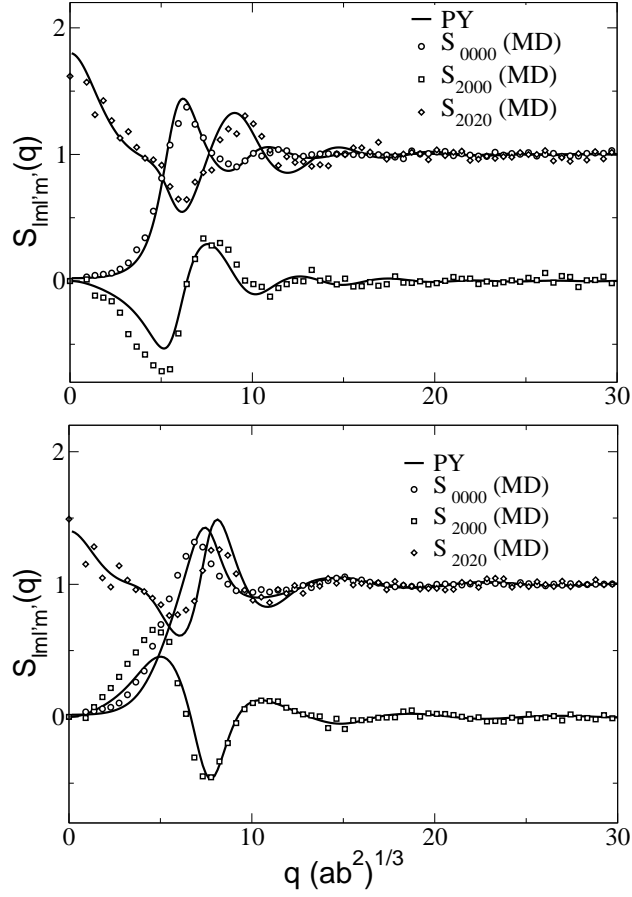

\includegraphics[width=0.5\textwidth]{PY-MD-Phi040EL040}
\vskip 0.2cm
\includegraphics[width=0.5\textwidth]{PY-MD-Phi040EL22}
\caption{$S_{lml'm'}\left({\bf q}\right)$ at $\phi=0.4$ for
$X_{0}=0.4$ (top) and $X_{0}=2.2$ (bottom).  Symbols are
simulation results, lines are PY predictions from
Ref.\protect\cite{letzlatz}.} \label{fig:diluted_very_oblate_prolate}
\end{figure}

Fig.~\ref{fig:diluted_very_oblate_prolate} shows data for the
tensorial correlation functions for values deviating stronger from
$X_0=1$ namely $X_0=0.4$ and $X_0=2.2$ at $\phi=0.4$. Compared to
the previous case, $S_{0000}$ is less structured, while the
opposite behavior is observed for both $S_{2000}$ and $S_{2020}$.
Even in this case, prolate and oblate ellipsoids are easily
distinguished from the $q$  dependence of $S_{2000}$: while for a
prolate ellipsoids  a maximum followed by a minimum is observed,
the opposite behavior characterizes  oblate ellipsoids. The
location of the first peak of  $S_{0000}$  is now shifted as
compared to the location of the $S_{2000}(q)$ extremum. It is also
interesting to observe that now a peak at $q=0$ is present in
$S_{2020}$,  signaling the build-up of a finite nematic
correlation length\cite{LetzSchilLatz} on approaching the nematic
transition. 

\begin{figure}
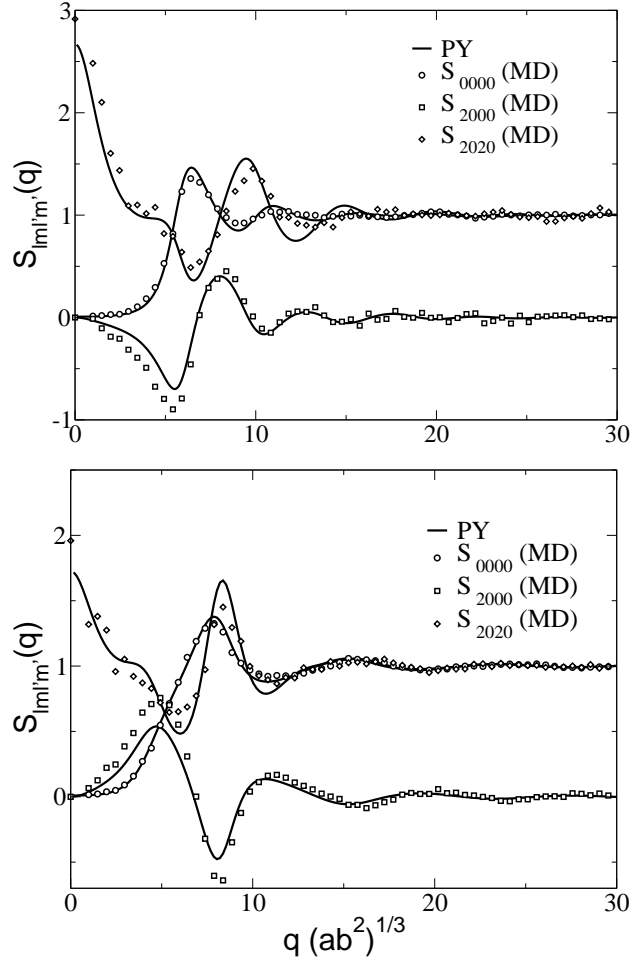

\includegraphics[width=0.5\textwidth]{PY-MD-Phi050EL040}
\vskip 0.3cm
\includegraphics[width=0.5\textwidth]{PY-MD-Phi050EL24}
\caption{$S_{lml'm'}\left({\bf q}\right)$ at values of $\phi$ near
the nematic transition line, for values of $X_{0}$ corresponding
to oblate/prolate ellipsoids; $X_0=0.4, $ $\phi=0.5$ (top) and
$X_0=2.4$, $\phi=0.5$ (bottom)}
\label{fig:near_MCT_oblate_prolate}
\end{figure}

\begin{figure}
\vskip 0.8cm
\includegraphics[width=0.5\textwidth,height=7.5cm]{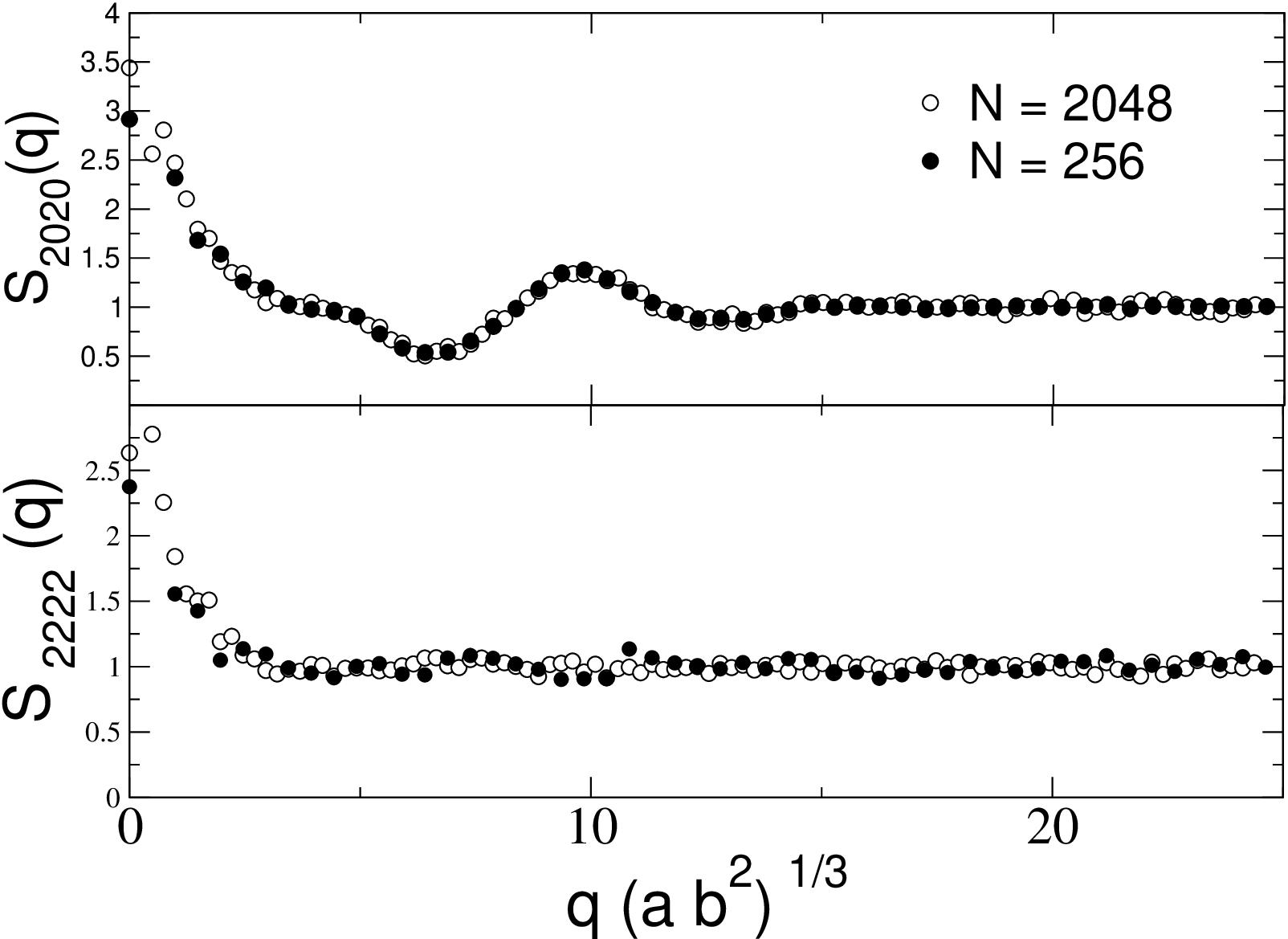}
\caption{Comparison of $S_{2020}\left({\bf q}\right)$ (top)
and  $S_{2222}\left({\bf q}\right)$ (bottom)
for $N=256$ (full circles) and $N=2048$ (open circles) ellipsoids at $\phi=0.50$, $X_{0}=0.40$.}
\label{fig:sq2m2m_N-N_oblate}
\end{figure}

\begin{figure}
\vskip 0.8cm
\includegraphics[ width=0.5\textwidth,height=7.5cm]{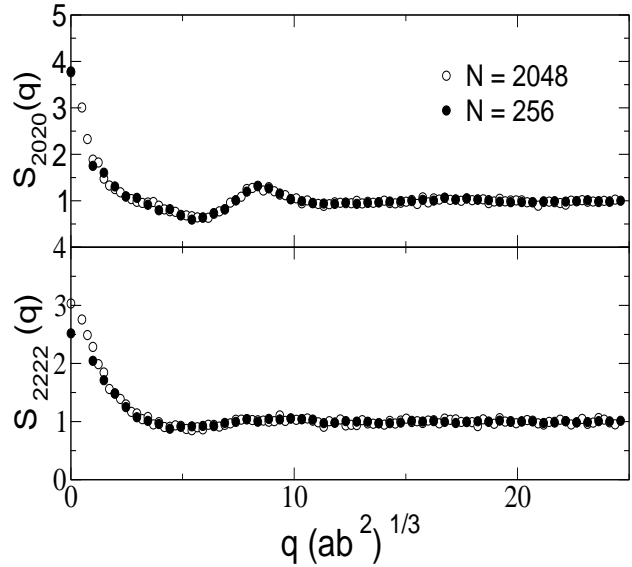}
\caption{ Comparison of $S_{2020}\left({\bf q}\right)$ (top)
and  $S_{2222}\left({\bf q}\right)$ (bottom)
for $N=256$ (full circles) and $N=2048$ (open circles) ellipsoids at $\phi=0.50$, $X_{0}=2.8$.}
\label{fig:sq2m2m_N-N_prolate}
\end{figure}

As a further case for comparing simulation results and theoretical
predictions, Fig.~\ref{fig:near_MCT_oblate_prolate} shows
$S_{lml'm}$ for $\phi=0.5$ and $X_0=0.4$ and $2.4$. These state
points are the closest points to the isotropic-nematic boundary
for which PY predictions are available from ref.\cite{letzlatz}.  Even in this case, the PY
results provide a satisfactory description of the $q$ dependence.
The growth of $S_{2020}\left(0\right)$ for $q\rightarrow0$ at
high/low elongations (Fig~\ref{fig:diluted_very_oblate_prolate} and Fig.~\ref{fig:near_MCT_oblate_prolate}) 
signals the presence of
a growing correlation length $\xi_{2020}$ for nematic order. If $\xi_{2020}$
would be of the order of the box-size, the results of our
simulations would be affected by finite size effects. To test for
the absence of such effects, we have simulated the two
representative points $\phi=0.50$, $X_{0}=0.40$ and $\phi=0.50$,
$X_{0}=2.80$ for $N=2048$ ellipsoids. As shown in
Fig.~\ref{fig:sq2m2m_N-N_oblate} and
Fig.~\ref{fig:sq2m2m_N-N_prolate}, the data for the biggest size
are consistent with the results for the smaller size at lower
$q$'s, providing evidence of the  absence of finite size effects
in the studied state point. Additionally,
Fig.~\ref{fig:sq2m2m_N-N_oblate} and
Fig.~\ref{fig:sq2m2m_N-N_prolate} depicts $S_{2222}(q)$ which has a
pronounced peak at $q=0$, and which is almost structureless for
larger $q$.
We have also calculated correlators with $l$ and/or $l'$ equal to $4$.
Those are not shown, because they are of less experimental relevance.
Nevertheless, we mention that they satisfactory agree with PY-result.

\subsection{Geometrical characterization of the $q$ dependence of
$S_{2000}$}
 
\begin{figure}
\includegraphics[width=0.25\textwidth]{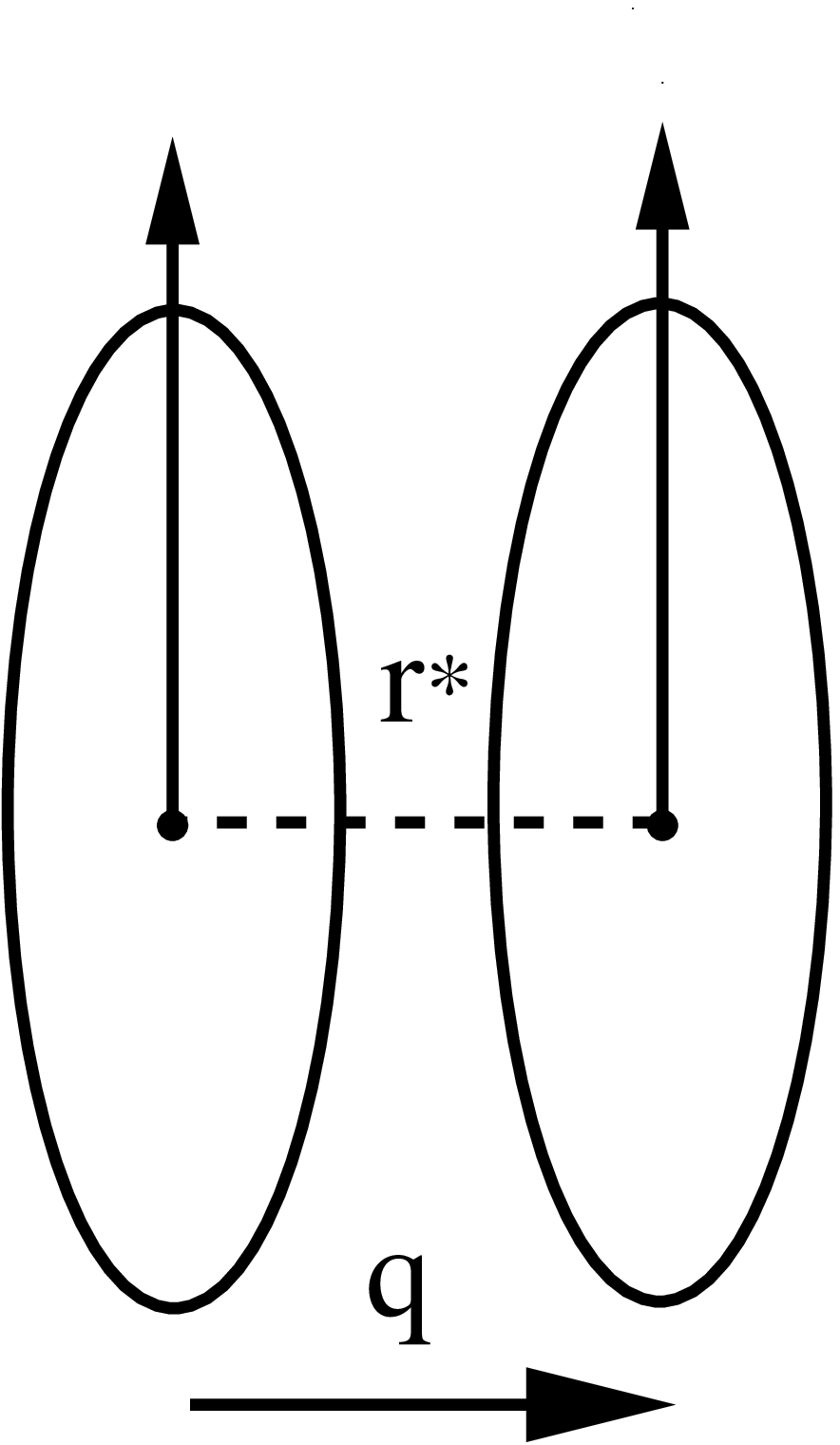}\textbf{(a)}
\vskip 0.5cm
\includegraphics[width=0.4\textwidth]{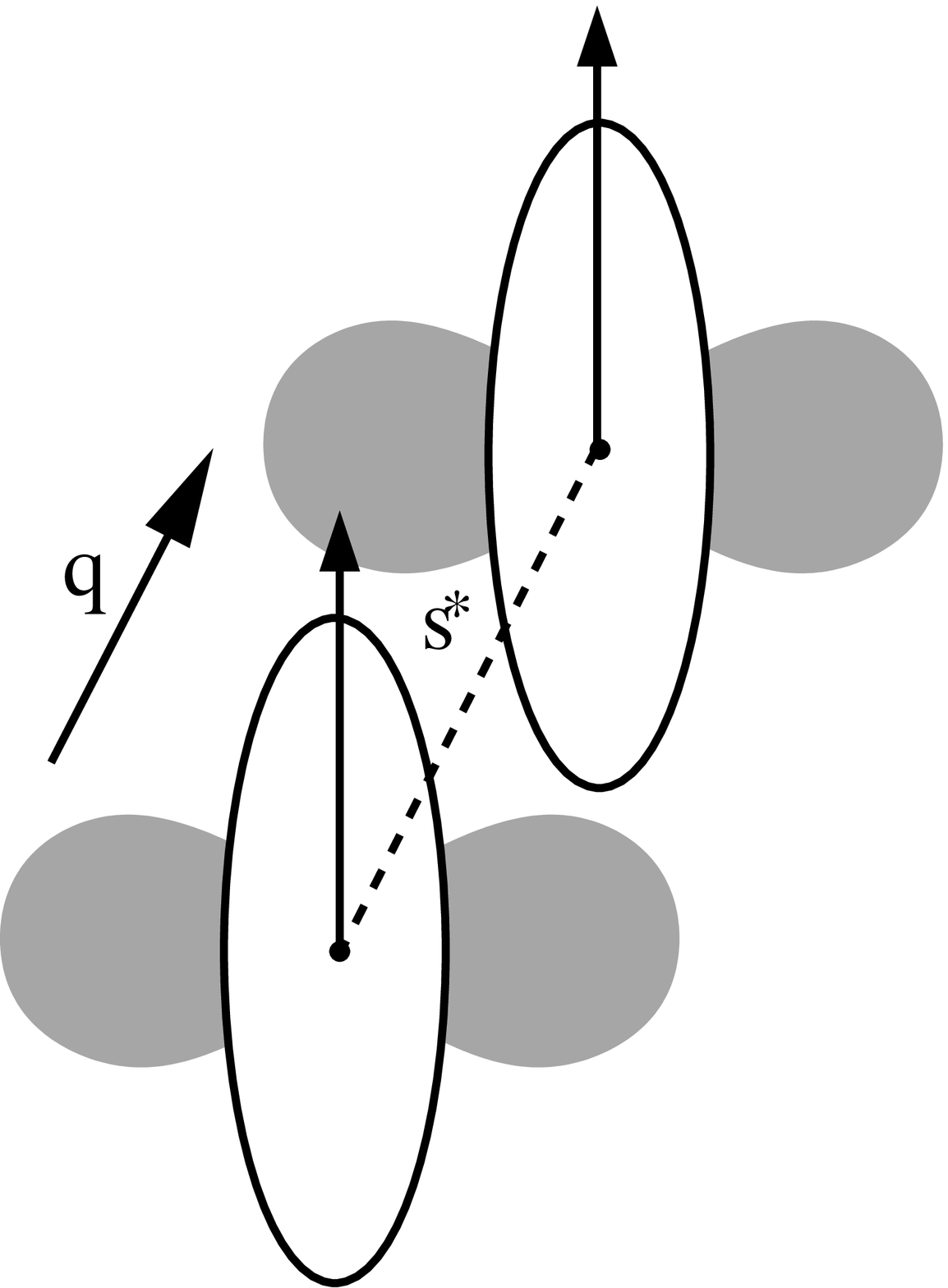}\textbf{(b)}
\caption{\label{fig:S2000_explain_prolate}
\textbf{(a)} Sketch of  configurations  of two prolate ellipsoids with a center-to-center distance 
close to $r^*$. The values of $Y_{20}$ in the $q$-frame is negative for the two prolate ellipsoids 
because their rotational symmetry axes are almost perpendicular to $q$.
\textbf{(b)} Configurations of ellipsoids with a center-to-center distance corresponding to $s^*$. 
The shaded region represents the volume excluded by particles at a distance $\sim r^*$.
The values of $Y_{20}$ in the $q$-frame is positive because the rotational symmetry axes of both ellipsoids 
are almost parallel to ${\bf q}$.
The lines with the arrows placed on the ellipsoids centers represent the axes of symmetry 
of the ellipsoids.
}
\end{figure}

An hand-waiving understanding of the anti-phase character of the
oblate/prolate $S_{2000}$ function at same $\phi$ can be obtained
by a geometrical analysis of the configurations. For the case of
prolate particles ($X_0>1$), the location $q^*$ of the first
minimum in $S_{2000}(q)$ is located at $q^* \approx 2 \pi/b$.  At
this small distance $r^* \approx b$, the two particles must be
almost parallel. 
Since $\bf q$ is along the $z$-axis ($q$-frame) and the relative distance 
of the two particles has to be parallel to $\bf q$ in order to give a non-zero 
contribution to $S_{2000}$, the polar angle $\theta$ of both ellipsoids is close to $\pi/2$.
This provides a negative contribution to $S_{2000}$ because 
$Y_{20}(\theta,\phi) \equiv (3 \cos^2\theta -1)/2 \approx -1/2 $
(Fig.~\ref{fig:S2000_explain_prolate}(a)) assumes its smallest possible
value. Hence, around $q^*$, where the majority of the pairs are
parallel, $S_{2000}$ will have a minimum.
On the other hand, at the first maximum of $S_{2000}$ ($q<q^*$) correlation between pairs of ellipsoids with relative distance $s^* \approx \max\{a,b\} > r^*$ is sampled. 
As we are working in the $q$-frame, the pairs of ellipsoids at a distance $s^*$ that will contribute to the maximum at $q^*$ are those whose relative distance is parallel to ${\bf q}$. If we take into account the excluded volume effects due to the ellipsoids at distance $\approx r^*$, we see that the sterically favoured configurations are the ones shown in Fig. \ref{fig:S2000_explain_prolate}(b). For such pairs of ellipsoids, the total contribution to $S_{2000}$ in the $q$-frame is  positive.
Therefore if we find an extremum at a distance $\approx s^*$ for prolate ellipsoids we expect that one to be positive.

\begin{figure}
\includegraphics[width=0.3\textwidth]{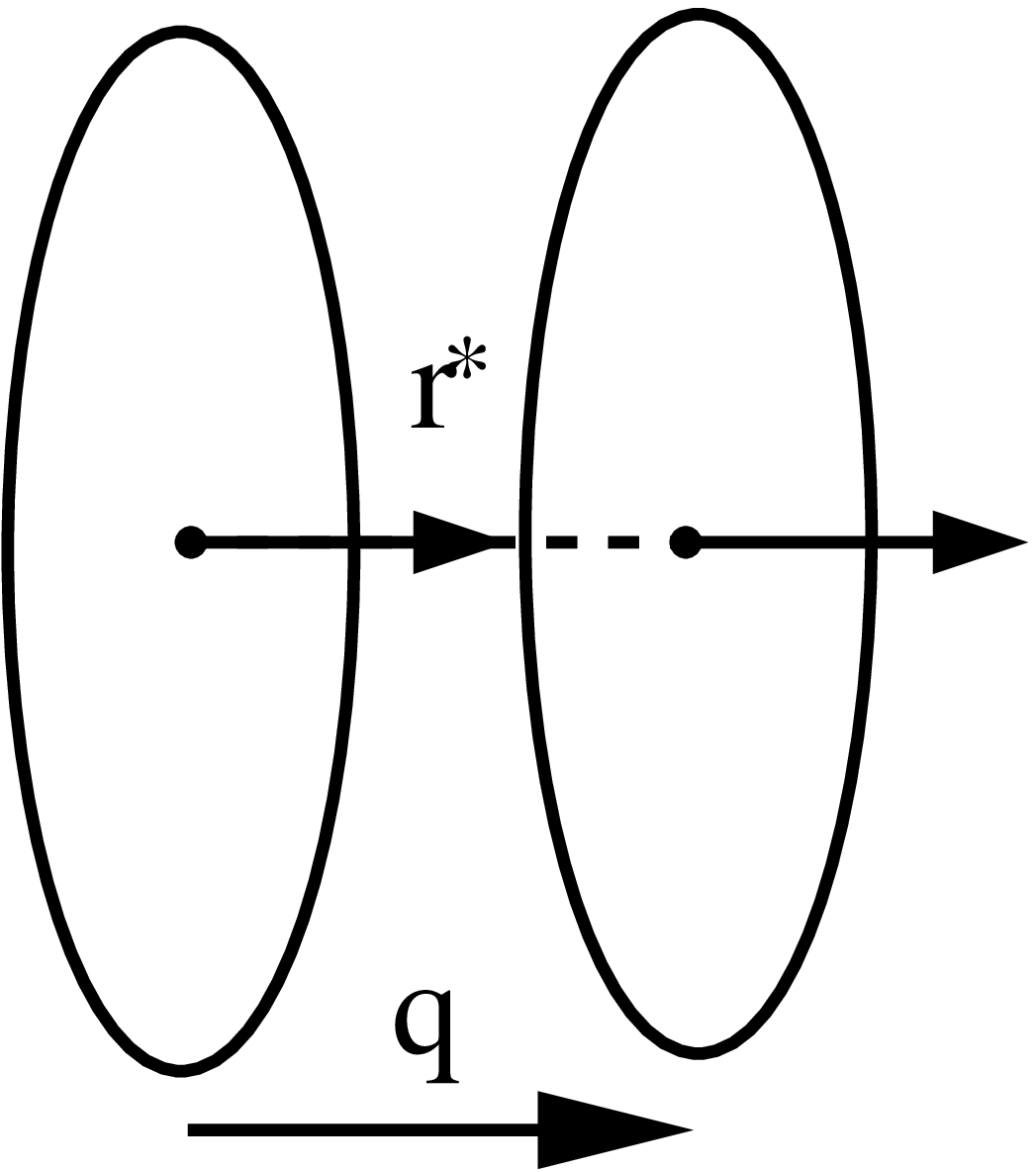}\textbf{(a)}
\vskip 0.5cm
\includegraphics[width=0.4\textwidth]{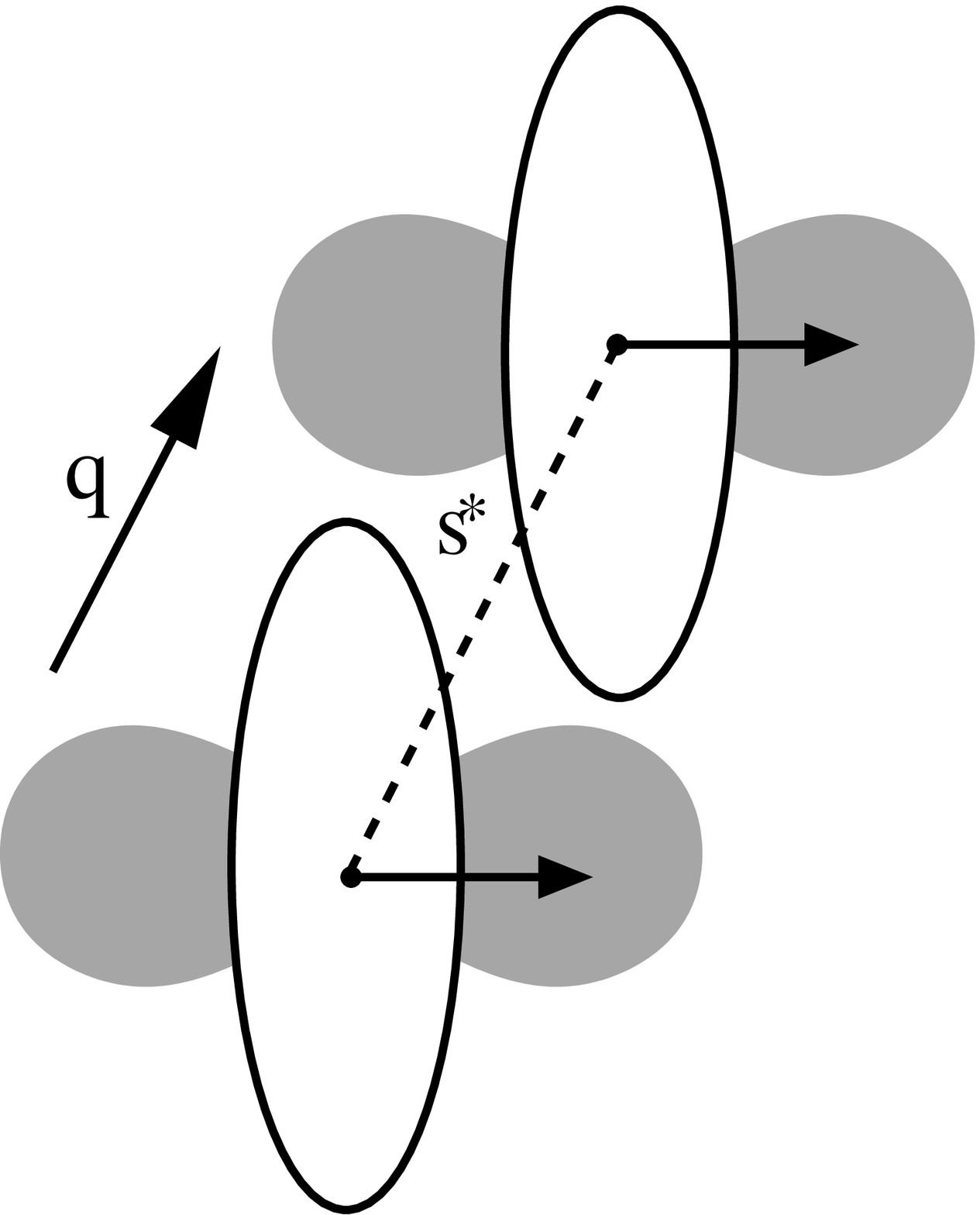}\textbf{(b)}
\caption{\label{fig:S2000_explain_oblate}
\textbf{(a)} Sketch of   configurations  of two oblate ellipsoids with a center-to-center distance 
close to $r^*$.  The values of $Y_{20}$ in the $q$-frame is positive for the two prolate ellipsoids because 
their rotational symmetry axes are almost parallel to ${\bf q}$.
\textbf{(b)} Configurations of ellipsoids with a center-to-center distance corresponding to $s^*$. 
The shaded region represents the volume excluded by particles at a distance $\sim r^*$.
The values of $Y_{20}$ in the $q$-frame is negative for prolate ellipsoids because the rotational symmetry
axes of both ellipsoids are almost perpendicular to ${\bf q}$.
The lines with the arrows placed on the ellipsoids centers represent the axes of symmetry 
of the ellipsoids.
}
\end{figure}

The analysis of the configurations contributing to the peaks of
$S_{2000}$ for oblate ellipsoids is analogous.
In the $q$-frame there is a shift of $\pm\pi/2$ in $\theta$.
Therefore, the sign of $Y_{20}$ is inverted and, consequently the
sign of $S_{2000}$ is inverted with respect to the prolate case
(see Fig.~\ref{fig:S2000_explain_oblate}).

\subsection{First order perturbation theory for $S_{lml'm'}$}
 
Specializing the results of Ref.~\cite{GrayGubbins} to hard
ellipsoids, it is possible to expand the radial distribution function
$g({\bf r}_{12}, {\bf u}_1, {\bf u}_2)$ in a power series in
$\varepsilon=X_0-1$:
 
\begin{equation} \label{eq1}
g({\bf r}_{12}, {\bf u}_1, {\bf u}_2) = {\displaystyle \sum_{i=0}^
\infty} g_i({\bf r}_{12}, {\bf u}_1, {\bf u}_2)
\end{equation}
 
\noindent where $g_n=O(\varepsilon^n)$, ${\bf r}_{12}$ is the
vector connecting two centers of the two ellipsoids and ${\bf
u}_i$ are the unit vectors along the rotational symmetry axes. The
first two terms of the expansion are
 
\begin{equation} \label{eq2}
g_0({\bf r}_{12}, {\bf u}_1, {\bf u}_2)=g_{HS} (r_{12})
\end{equation}
 
\begin{eqnarray} \label{eq3}
&& g_1({\bf r}_{12}, {\bf u}_1, {\bf u}_2)= \nonumber\\
&&=-\bar{\sigma} g_{HS} (r_{12})
\gamma ({\bf u}_1, {\bf u}_2, {\bf e}_{12}) \, \delta (r_{12}-\bar
\sigma
)\nonumber\\
&&-\rho \bar{\sigma}  \, \int d^3 r_3 \, g_{HS} ({\bf
r}_{12}, {\bf r}_{13}, {\bf r}_{23} )
\cdot \nonumber\\
&& \cdot [\langle \gamma({\bf u}_1, {\bf u}_3; {\bf
e}_{13})\rangle_{{\bf u}_3} \delta (r_{13} - \bar{\sigma}) + \\
&& + \langle \gamma ({\bf u}_2, {\bf u}_3, {\bf e}_{23} )
\rangle_{{\bf u}_3}
\, \delta (r_{23} -\bar{\sigma}) ] \quad
\end{eqnarray}
 
\noindent were $g_{HS} (r_{12})$ and $g_{HS} ({\bf r}_{12}, {\bf
r}_{13}, {\bf r}_{23})$ are the static 2-particle and 3-particle
distribution function of hard-spheres with effective diameter
$\bar{\sigma}$ depending on $\varepsilon$, ${\bf r}_{ij} = r_{ij}
{\bf e}_{ij}$, $\rho=N/V$ is the number density and
 
\begin{equation} \label{eq4}
\langle f ({\bf u})\rangle _{\bf u} \equiv \frac{1}{ 4 \pi}
\,\int\limits^{2 \pi}_0 d \varphi \, \int\limits^{\pi}_0 d
\vartheta \sin \vartheta \, f({\bf u} (\vartheta, \varphi)) \quad
.
\end{equation}
 
\noindent The effective diameter $\bar{\sigma}$ is defined as
$\bar{\sigma} =\langle \langle d({\bf u}_1,{\bf u}_2,{\bf e}_{12}
) \rangle_{{\bf u}_1} \rangle_{ {\bf u}_2}$, where $d({\bf
u}_1,{\bf u}_2,{\bf e}_{12})$ is the distance at contact of two
ellipsoids with axis along ${\bf u}_i$ and ${\bf e}_{12}={\bf
r}_{12}/r_{12}$ is the direction between the centers of the
ellipsoids. The function $\gamma$ measures the "non-sphericity" of
the potential and is defined self-consistently as $d({\bf
u}_1,{\bf u}_2,{\bf e}_{12}) =\bar{\sigma} \, \left[ 1 +\gamma
({\bf u}_1, {\bf u}_2, {\bf e}_{12}) \right]$ . At first order in
$\varepsilon$, $\gamma$ and $\bar{\sigma}$ can be evaluated
analytically from geometrical considerations
 
\begin{equation} \label{eq7}
\bar{\sigma}  = b [1+ \frac{\varepsilon}{3} + O(\varepsilon^2)]
\end{equation}
 
\begin{equation} \label{eq8}
\gamma ({\bf u}_1, {\bf u}_2, {\bf e}_{12}) =
\frac{\varepsilon}{3}  [
P_2 ({\bf e}_{12}\cdot {\bf u}_1) + P_2 ({\bf e}_{12} \cdot {\bf
u}_2 ) ] + O(\varepsilon^2) \quad.
\end{equation}
 
\noindent
where $P_l$ is the Legendre polynomial of order $l$.

The general expression of the molecular structure factor $S_{lml'm}$
in terms of $g({\bf r}_{12}, {\bf u}_1, {\bf u}_2)$ is
 
\begin{eqnarray} \label{eq11}
&&S_{lml'm} ({\bf q}) = \delta_{ll'} +i^{l'-l} \rho\,4\pi \int d^3
r_{12} \langle \, \langle \, g ({\bf r}_{12}, {\bf u}_1, {\bf u}_2)
\nonumber\\
&&\cdot e^{i {\bf q}\cdot{\bf r}_{12}} Y_{lm}^* ({\bf u}_1)
Y_{l'm} ({\bf u}_2) \, \rangle_{{\bf u}_1} \, \rangle_{{\bf u}_2}
\end{eqnarray}

Using the properties of Legendre polynomials, we have that
$\langle \gamma ({\bf u}, {\bf w}, {\bf e} ) \rangle_{\bf
w}=(\varepsilon/3)P_2 ({\bf e}\cdot {\bf u})$, so we can rewrite
$g$ up to first order:
 
 
\begin{eqnarray}
g(\textbf{r}_{12},\textbf{u}_1,\textbf{u}_2) &=&
g_{HS}(\textbf{r}_{12})
-\frac{\epsilon\bar{\sigma}}{3}  \left\{ g_{HS}(\bar{\sigma}) \cdot
\right . \left[P_{2}(\textbf{u}_1\cdot\textbf{e}_{12})\right .
+\cr &&\left
.+P_{2}(\textbf{u}_2\cdot\textbf{e}_{12})\right]\,
\delta(r_{12}-\bar{\sigma})+\cr
&& +\rho \, \int d^{3}r_{3}\, 
g_{HS}(\textbf{r}_{12},\textbf{r}_{23},\textbf{r}_{13}) \cdot\cr &&
\left[ \delta(r_{13}-\bar{\sigma})\, P_{2}\right
.\left(\textbf{u}_1\cdot\textbf{e}_{13}\right)+\cr &&\left
.\delta(r_{23}-\bar{\sigma})\,
P_{2}\left(\textbf{u}_2\cdot\textbf{e}_{23}\right) \right]\left.\right\}
\end{eqnarray}
 
The zero-th order  $g_0$ is spherically symmetric and will
contribute with a diagonal term $S^0_{lmlm}$. The first order term
contains functions of the form $P_2({\bf u}_i\cdot {\bf e}_{ij})$
that can be recasted in terms of linear combinations of spherical
harmonics $Y_{lm}({\bf u}_i)$ with $l=0,2$; so the first order
term will contribute only to $S_{lml'm}$ with $l,l'=0,2$ and $l
\neq l'$.
 
In particular, in the $q$-frame the $\rho$-independent part of
$g_1$ contains only linear combinations of $Y_{00}$ and $Y_{20}$; it
is then possible to evaluate $S_{lml'm}$ 
with the result
 
\begin{eqnarray}
S_{lml'm}(q,\epsilon) & = & \delta_{ll'} S^0_{lmlm}(q)  - \epsilon
\frac{2\pi}{3\sqrt{5}}\rho
\bar{\sigma}^{3}g_{HS}\left(\bar{\sigma}\right)\cdot\cr
 && \cdot\left[\left(x^{-1}-2x^{-3}\right)\sin x+2x^{-2}\cos x
\right]\cr
 && \cdot
\left[\delta_{lm,20}\delta_{l'm,00}+\delta_{lm,00}\delta_{l'm,20}\right]\cr
 &&+ \rho\varepsilon
 {\cal F}_{lml'm}(\bar\sigma) + O(\varepsilon^2)
\end{eqnarray}
\noindent where $x=\left|q\right|\bar{\sigma}$ and ${\cal
F}_{lml'm}(\bar\sigma)$ is a function of $\bar\sigma$ that does
not depend on $\varepsilon$. Its calculation requires the
knowledge of the static three-point correlator $g_{\rm HS}
(\vec{r}_{\rm 12}, \vec{r}_{\rm 13}, \vec{r}_{\rm 23})$, which,
however, is not known exactly.

\begin{figure}
\includegraphics[ width=0.5\textwidth]{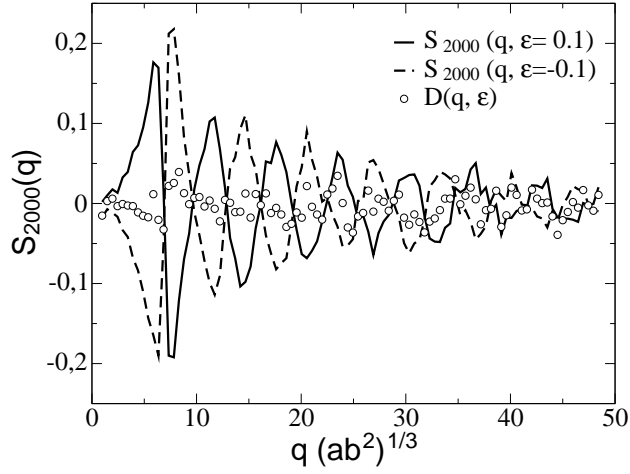}
\vskip 1.0cm
\includegraphics[ width=0.48\textwidth]{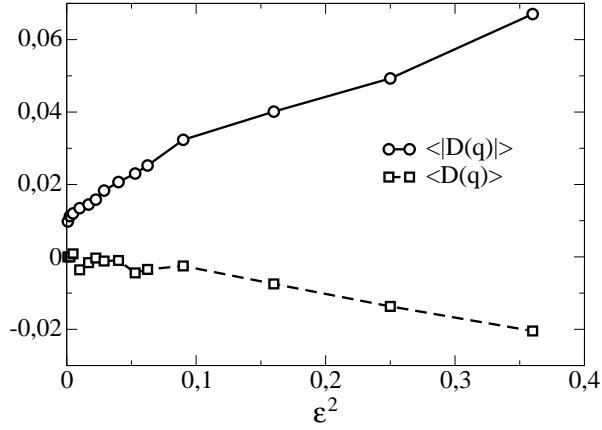}
\vskip 0.5cm
\vskip 0.5cm
\caption{Check of Eq.\protect\ref{eq:prediction} for the case $\phi=0.49$.
The upper panel shows $S_{2000}(q)$ for $X_0=1\pm\epsilon$ and $D(q,\epsilon)$ for $\epsilon=0.10$.
The lower panel shows the average of $|D(q,\epsilon)|$ (empty circles)  over all $q<50$ from the simulation as a function of $\epsilon^2$ for several values of $\epsilon$. The behavior at low $\epsilon$ is linear in $\epsilon^2$ as predicted from first order perturbation theory. The presence of a constant term in $<|D(q,\epsilon)|>$ for $\epsilon\rightarrow0$ is due to the noise implicit in the measures. In fact, the values $<D(q,\epsilon)>$ (empty squares) at small $\epsilon$ are dominated by the noise and are scattered around zero. The $\epsilon^2$ regime seems to break down around $\epsilon\sim0.3$
%
}

\label{fig:eps_check}
\end{figure}

 
Therefore, $S_{2000}$ shows a peak of opposite sign around $q \sim
{\bar{\sigma}}^{-1}$ as observed above when discussing the
simulation data; in general, first order theory predicts
\begin{equation}
D(q,\epsilon)=S_{2000}(q,\epsilon)+S_{2000}(q,-\epsilon)=0
\label{eq:prediction}
\end{equation}
for $\varepsilon$ small enough.
We check this last property by our simulations.
Fig.~\ref{fig:eps_check}-(top) shows $S_{2000}(q,\epsilon)$,
$S_{2000}(q,-\epsilon)$ and $D(q,\epsilon)$ for $\varepsilon = 0.05$.
The sum $D(q,\epsilon)$ vanishes within the error, supporting  the
prediction of Eq.~\ref{eq:prediction}.
To gather a feeling of the range of validity
of this prediction we show in  Fig.~\ref{fig:eps_check}-(bottom)  the average value (over all $q < 50$) 
of $|D(q,\epsilon)|$ and $D(q,\epsilon)$ as a function of $\epsilon^2$.
The behavior of $<D(q,\epsilon)>$ and $<|D(q,\epsilon)|>$ confirms the linear dependence 
on $\epsilon^2$ predicted from Eq.~\ref{eq:prediction} for $\epsilon \lesssim 0.1$.

\section{Conclusions}
 
In this paper we have presented a comparison between PY predictions and
MD simulation results  for the tensorial correlators  in a hard ellipsoids liquid, close to the
phase coexistence lines. 

The major focus has been on the calculation of static molecular correlation functions.
In contrast to earlier work we have not chosen the $r$-dependent coefficients $g_{ll'm}
(r)$ of an expansion with respect to rotational invariants but the
tensorial correlators $S_{lm l'm'} ({\bf q})$ in $q$-space. Those
have the advantage that they can be obtained from scattering
experiments, at least for $l,l' \le 2$. The comparison of the correlators
from MD-simulations with the corresponding ones
from PY-theory \cite{letzlatz} is rather satisfactory for all correlators and
all pairs of ($X_0, \phi)$ we have studied. Accordingly, the good
agreement between results from PY-theory and an earlier
MD-simulation found in Ref. \cite{A14} is confirmed. An
interesting observation made is the qualitative difference of
the nondiagonal correlator $S_{\rm 2000} (q)$ for oblate and
prolate shape. Since the back transform to real space is a linear
procedure, this qualitative different behavior should also exist
for $G_{\rm 2000} (r)$ which is related to the coefficient $g_{\rm
200} (r)$. Indeed, Fig.~1a for $X_0=2$ and Fig.~4 for $X_0=1/3$
from Ref. \cite{A11} show that the first extremum of $g_{\rm 200}
(r)$ is a minimum and a maximum, respectively.

The qualitative shape dependence of $S_{\rm 2000} (q)$ has been
proven analytically. Using first order perturbation theory with
respect to $\varepsilon=X_0-1$ we have shown that\\

$S_{\rm 2000} (q) \mid_{\rm oblate} \approx - S_{\rm 2000} (q)
\mid_{\rm oblate} \quad . $\\

We have not attempted to compare $S_{\rm 2000} (q)$ from this
perturbational approach, with the corresponding result from our
MD-simulation and PY-theory, because one needs the static three-point
correlator for hard spheres as an input which is not known.

It has recently been predicted \cite{A17} that the time-dependent
correlator $S_{\rm 2000}({\bf q},t)$, which is a measure of the
coupling between the center of mass ($l=0$) and orientational
(``quadrupolar'' part $l=2$) motion, has an effect on the
light scattering spectra. This effect has been found experimentally
\cite{A18} and might offer the possibility to check how far
spectra from light scattering experiments may allow to discriminate
between oblate and prolate particles.

Finally, we have checked the growth of nematic order. In contrast
to the PY-result in Ref. \cite{A10} the authors of Ref. \cite{letzlatz} have
found that $S_{\rm 2020} (q=0)$ from PY-theory diverges at a
critical volume fraction $\phi_c(X_0)$ for $X_0 \gtrsim 2$ and
$X_0 \lesssim 0.5$. At $\phi_c$ an isotropic-nematic transition
occurs. Our simulation reproduces the shape and the growth of the
peak of $S_{\rm 2020} (q)$ at low $q$'s. We have also demonstrated
that, for the chosen values of $X_0$ and $\phi$,  no finite size effects influences this peak. 
We have not attempted to determine $\phi_c(X_0)$ from our simulation since this type of
analysis would request much larger system sizes than we are currently able to
simulate. 

The PY tensorial correlators have been recently 
used as input in molecular mode-coupling theory to evaluate the
glass lines in the $(X_0-\phi)$ plane.  The theoretical calculations suggest the possibility
of a new mechanism of slowing down of the dynamics driven by the
increase of the nematic order. Therefore, the validity of the PY predictions, 
particularly for the peak of $S_{2020}$ close to $q=0$,
presented in this work confirms that this new mechanism is not arising from a failure of the PY predictions, 
but it is a genuine prediction of  the  molecular mode-coupling  approach.


\begin{acknowledgments}
 
We acknowledge support from MIUR-Firb. We also thank Martin Letz
for providing the PY-results. R.S. gratefully acknowledges the
hospitality at the Dipartimento di Fisica of Universit\`a di Roma
``La Sapienza'' where this has been started.
\end{acknowledgments}

\bibliography{CompareMDPY.bib}
\newpage
 
%
%

\end{document}